\documentclass[conference, a4paper]{IEEEtran}
\IEEEoverridecommandlockouts
\usepackage{cite}
\usepackage{amsmath,amssymb,amsfonts}
\usepackage{graphicx}
\usepackage{textcomp}
\usepackage{xcolor}
\def\BibTeX{{\rm B\kern-.05em{\sc i\kern-.025em b}\kern-.08em
		T\kern-.1667em\lower.7ex\hbox{E}\kern-.125emX}}
\usepackage{array}
\usepackage{mdwmath}
\usepackage{mdwtab}
\usepackage{caption}
\usepackage{amsfonts}
\usepackage{url}
\usepackage{balance}
\usepackage{algorithm}
\usepackage{algorithmic}
\usepackage{lmodern}
\usepackage[tight,footnotesize]{subfigure}
\usepackage[caption=false,font=footnotesize]{subfig}
\usepackage[utf8]{inputenc}
\usepackage[left=1.62cm,right=1.62cm,top=1.9cm]{geometry}
\begin{document}

	\title{More Capacity from Less Spectrum: Tapping into Optical-layer Intelligence in Optical Computing-Communication Integrated Network \\
		
	}
	
	\author{\IEEEauthorblockN{Dao Thanh Hai}
		\IEEEauthorblockA{\textit{RMIT University Vietnam} \\
			hai.dao5@rmit.edu.vn}
		\and
		\IEEEauthorblockN{Shuo Li}
		\IEEEauthorblockA{\textit{RMIT University} \\
			shuo.li2@rmit.edu.au}
		\and
		\IEEEauthorblockN{Isaac Woungang}
		\IEEEauthorblockA{\textit{Toronto Metropolitan University} \\
			iwoungan@torontomu.ca}
				
	}
	
	\maketitle
	\begin{abstract}
	
		Optical fiber communications and networks constituting the backbone of Internet infrastructure have been continuing evolving technologically and architecturally to meet the explosive traffic growth. On the technological front, recent key advancements including spectrally and spatially flexible transmission, and wide-band optical transmission hold the promise of order-of-magnitude capacity gain. Concurrently, architectural innovations seek to capitalize on this expanded capacity to enable more scalable, cost-effective, and energy-efficient data transport. A prominent example is the optical-bypass architecture, which has now become widely deployed since the 2000s due to its substantial operational and capital cost savings over the predecessor optical-electrical-optical (O-E-O) mode. Driven by massive investments and consequently significant progresses in optical computing and all-optical signal processing technologies lately, this paper presents a new architectural paradigm for next-generation optical transport network, entitled \textit{optical computing-communication integrated network}, which is capable of providing dual services at the optical layer, namely, computing and communication. This approach seeks to exploit the potential for performing optical computing operations among lightpaths that traverse the same intermediate node. \textit{Optical-layer intelligence concept} is thus introduced as the capability to perform computing / processing at the lightpath scale to achieve greater spectral and/or computing efficiency. A case study focusing on optical aggregation operation is introduced, highlighting the key differences between optical computing-communication integrated network and its current counterpart, optical-bypass ones. A mathematical formulation for optimal designs of optical-aggregation-enabled network is then provided and performance comparison with traditional optical-bypass model is drawn on the realistic NSFNET topology.

	\end{abstract}
	
	\section{Introduction}
	Broadband Internet services have become essential in modern life, driving continuous and rapid growth in Internet traffic\textemdash estimated at a compound annual growth rate (CAGR) of $24\%$ between 2021 and 2026 \cite{Cisco22}. This surge has been largely supported by the robust infrastructure of optical communication networks thanks to several billion kilometers of optical fiber being installed around the globe today. In response to the continually explosive traffic growth year-by-year, network operators worldwide have been innovating their fiber-optic infrastructure technologically and architecturally, aiming at greater spectral efficiency, thus expanding the network capacity. Indeed, since the introduction of the first fiber-optic transmission links in the 1970s offering only a few Mb/s capacity, the per-fiber throughput has been increasing roughly five order-of-magnitude, now entering the Pb/s era, with optical systems scaling across multiple spatial dimensions\textemdash from ultra-short links spanning just tens of meters to a few kilometers within data centers, to vast long-haul networks stretching across thousands of kilometers over continents \cite{OTN}. Nevertheless, the quest for more spectral efficiency relying on the higher modulation formats and/or higher bit-rate per channel appears to reach the Shannon's limit, thus necessitating new strategies for gaining more capacity. In this context, it is important to note that, beyond simply expanding the network capacity, network operators also prioritize efficient utilization of this ever-growing capacity so that more demands could be effectively accommodated within a given spectrum resources, or alternatively achieving more capacity at less spectrum cost. To achieve this, architectural innovations have been developed\textemdash not by solely increasing raw capacity, but by reducing effective traffic loads and optimizing resource usage, thereby lowering the capital and operational expenditure per transmitted bit \cite{all-optical}. \\
	
	From an architectural standpoint, optical networking has seen a significant shift since the early 2000s\textemdash from the costly optical-electrical-optical (O-E-O) mode to the highly cost-efficient and energy-efficient optical-bypass configuration. Unlike O-E-O mode where transiting lightpaths undergo repeated conversions between optical and electrical domains, optical-bypass operation enables direct optical cross-connection, eliminating unnecessary conversions and therefore significantly boosting efficiency \cite{efficient, Simmons}. Over the past two decades, this paradigm has evolved from a theoretical concept into a mainstream deployment adopted by the majority of network carriers worldwide \cite{all-optical}. However, despite its successes, the scalability of this network architecture remains challenged by the reliance on electrical domain at end nodes for signal processing and/or computing purposes, particularly in the context of emerging large AI models growing at an exponential rate, and yet demanding minimal environmental impact \cite{AI_optics5}. \\
	
	In optical-bypass networks, node-level functions typically include signal add/drop and lightpath cross-connect. However, when multiple lightpaths traverse to a common intermediate node, they must be isolated from each other in either time, frequency, or spatial domain to avoid unwanted interference \cite{nodearchitecture}. This reveals a fundamental limitation, as a range of optical signal processing and/or computing operations could be applied between these transitional lightpaths to produce output signals that are either potentially more spectrum-efficient than the original inputs or to serve computing purposes. Driven by massive investments and consequently significant progresses in optical computing and all-optical signal processing technologies lately \cite{roadmap2024, nature, nature2}, we propose a new architectural paradigm for future optical networks, namely, \textit{optical computing-communication integrated network} where both computing and communication services could be provided at the optical layer, reshaping the boundaries between communication and computation. Our proposal is essentially defined by the added function of optical nodes permitting optical computing between transitional lightpaths to reduce the effective network traffic load. We thus coin the term, \textit{optical-layer intelligence}, to highlight the new capability of optical nodes in performing computing at the speed of light between lightpaths to achieve greater spectral and/or computing efficiency. This paper is a continuation of our recent efforts in investigating a new optical network architecture that is capable of providing dual services, computing and communication, at the optical layer \cite{hai_comcom2, hai_wrap, hai_tnsm, hai_springer3, hai_oft24, hai_AINA, hai_apnet, hai_IoT1, hai_IoT2}. \\ 
	
	The paper is organized as follows. Section II presents an illustrative example centered on optical aggregation, emphasizing the key distinctions between the proposed optical computing-communication integrated architecture and traditional optical-bypass networks. Section III introduces the mathematical framework for optimizing the design of optical-aggregation-capable networks. Section IV reports simulation results using the realistic NSFNET topology to evaluate the spectral efficiency achieved through aggregation-aware routing compared to conventional approaches. Finally, Section V offers concluding remarks and discusses future research directions. \\

	\section{Optical-layer Intelligence in Optical Computing-Communication Integrated Network: The case of Optical Aggregation Operation}
	
	Resurgences of interests in optical computing have been gaining momentum in recent years thanks to massive investments centering around special-purpose computing operations, in the shadow of the end of Moore's law and the AI-driven computing demands \cite{nature4, nature5}. Most notably and in a recent keynote of NVIDIA CEO, the company has made a bold commitment to photonic computing by integrating co-packaged silicon photonics into its next-generation networking switches, aiming to overcome bandwidth and energy bottlenecks in AI-scale data centers and signaling a long-term shift toward light-speed, scalable infrastructure \cite{huang2025photonic}. While many optical computing operations have then been developed, in this paper we focus on optical aggregation and de-aggregation that could have the potential to revolutionize future optical transport networks \cite{optical_processing_1, optical_processing_2}. \\ 
	
	Recent remarkable advancements in optical aggregation and de-aggregation technologies have enabled the direct combination of multiple lower bit-rate channels into a single higher-capacity stream of higher-order modulation format within the optical domain \cite{aggregation1, aggregation2}. As these technologies continue to mature, they present valuable opportunities to rethink the traditional role of optical switching nodes in optical-bypass networks. Notably, equipping optical nodes with all-optical aggregation capabilities challenges the long-standing assumption that transiting lightpaths must remain isolated to prevent unwanted interference, thus heralding a paradigm shift that allows computing operations at the lightpath scale and unlocking new potential for improving spectral efficiency \cite{hai_tnsm, hai_apnet, hai_IoT1}. \\
	
	Consider two traffic demands requesting 400 Gbps originating from node $A$ and node $B$, respectively to a common node $C$. In an optical-bypass network, provisioning these demands involves determining suitable routes and assigning wavelengths independently for two lightpaths. As illustrated in Fig. 1, both lightpaths traverse the same links, $XI$ and $IC$, and are optically switched at the intermediate node $X$. Consequently, two distinct wavelengths are required across each shared link, resulting in a total cost of two wavelength count and six wavelength-link units across the network. In the new paradigm of optical computing-communication integrated network, optical nodes equipping with optical-layer computing capabilities open up new possibilities for in-network computation which has the potential for achieving greater capacity efficiency. One illustrative scenario involves enabling all-optical aggregation at intermediate node $X$. In this setup, the two 400G QPSK signals transmitted over the wavelength $\lambda_1$ are combined into a single 800G 16-QAM signal on the same wavelength $\lambda_1$. This aggregation, which merges multiple lower-rate channels into a single higher-capacity stream using a higher-order modulation format, significantly enhances spectrum efficiency. Upon reaching the shared destination node $C$, the aggregated signal can be optically de-aggregated back into its original components, as depicted in Fig. 2. In this configuration, only one wavelength, that is, $\lambda_1$, is required for the whole network as three lightpaths, from $A$ to $X$, $B$ to $X$ and $X$ to $C$, do not share a same link and the total spectral cost amounts to just four wavelength-link units, demonstrating significant spectral savings. Note that the computing sense for the optical aggregation operation could be interpreted as the summation of bits per symbol from two QPSK lightpaths of 2 bits/symbol to a new 16-QAM lightpath of 4 bits/symbol (Fig. 2). 
	
	\begin{figure}[!ht]
		\centering
		\includegraphics[width=\linewidth, height = 3.7cm]{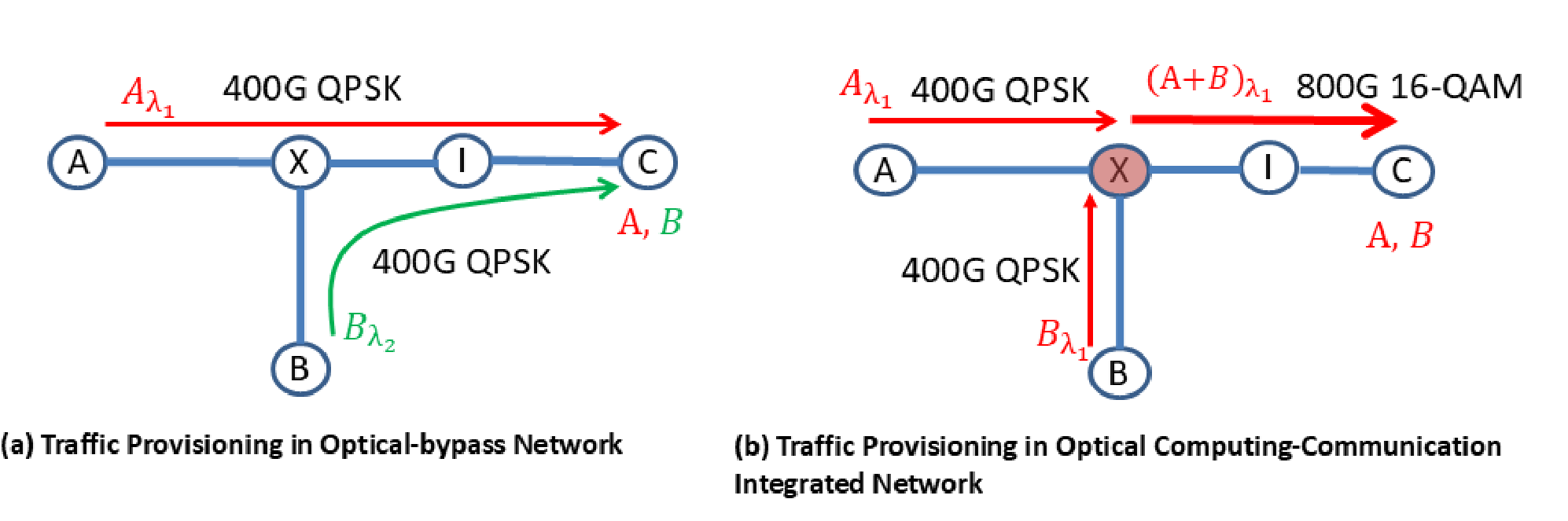}
		\caption{Traffic Provisioning in Optical-bypass Networking vs. Optical Computing-Communication Integrated Network}
		\label{fig:i3}
	\end{figure}
	
	\begin{figure}[!ht]
		\centering
		\includegraphics[width=\linewidth, height = 4.7cm]{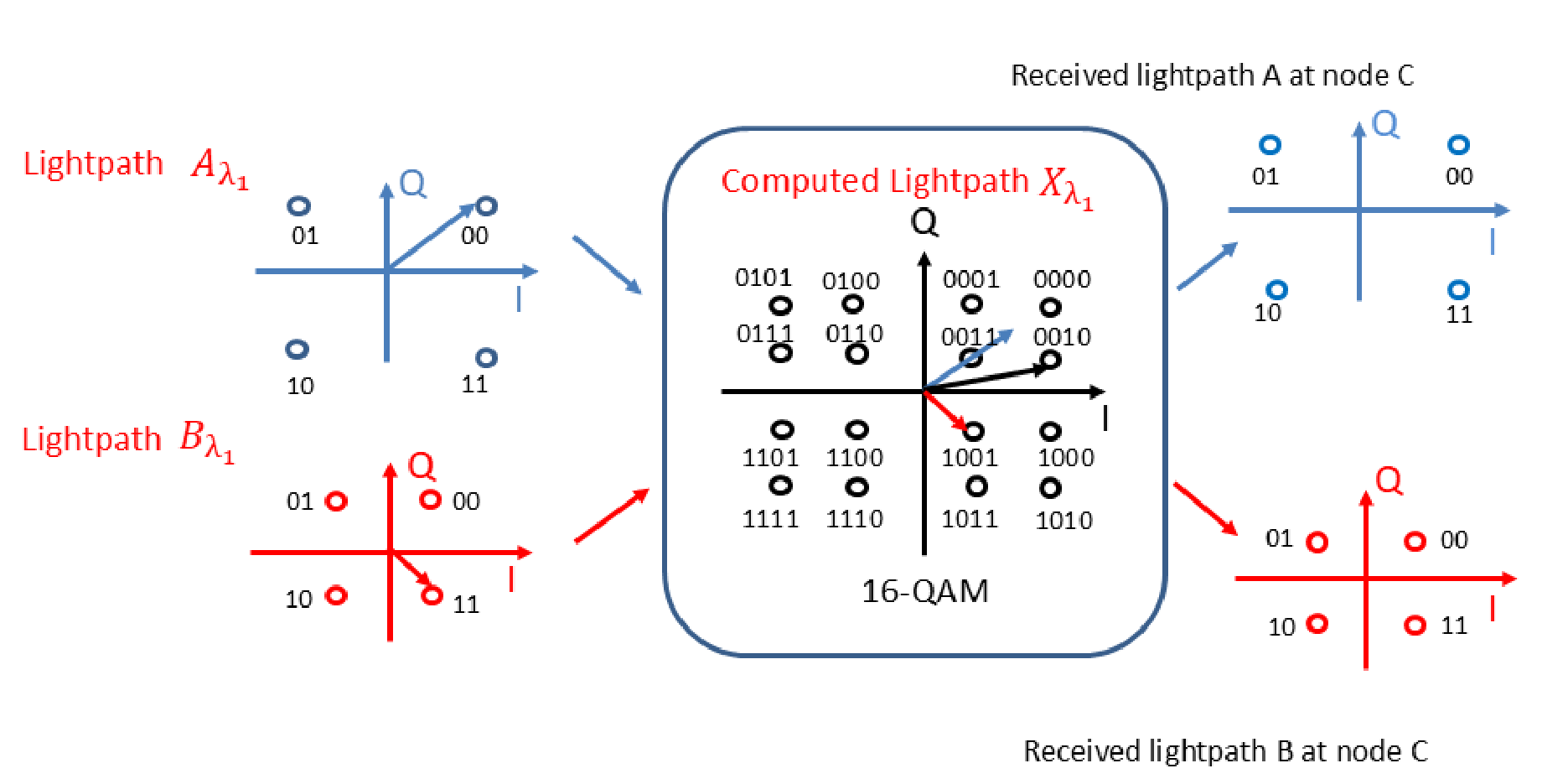}
		\caption{Schematic Diagram of Optical Aggregation Operation at Node X and De-aggregation at Node C}
		\label{fig:i4}
	\end{figure}
	
	The above example highlights the effectiveness of optical aggregation and de-aggregation in improving spectral efficiency within optical transport networks. The optical computing-communication integrated network model introduces a new level of architectural flexibility by allowing controlled interference between multiple transitional lightpaths\textemdash an innovation that carries significant implications for network design strategies aiming to fully harness this potential. Specifically, in the context of optical aggregation, key design challenges include identifying suitable demand pairs for aggregation, selecting optimal aggregation node locations, and determining routing paths for the resulting aggregated lightpaths. The following section presents a mathematical framework developed to address these challenges systematically.

	\section{A Mathematical Formulation for Optimizing Wavelength Link Utilization in Optical-aggregation-enabled Network}
	
	To integrate optical aggregation and de-aggregation into network design and planning, we focus on a practical scenario where aggregation occurs only between two lightpaths sharing the same line rate and modulation format, and de-aggregation is limited to their common destination node. This setup is feasible with current optical aggregators capable of combining two QPSK signals into a single 16-QAM channel \cite{optical_processing_2}. \\
	
	The spectral efficiency gain from such operations depends on the network’s physical topology, traffic patterns, routing schemes, and the location of aggregation nodes. To quantify this, we propose a mathematical model that minimizes wavelength link cost\textemdash or equivalently, maximizes aggregation benefits. Given the physical topology and demand matrix, the model identifies optimal routing paths, suitable demand pairs for aggregation, and corresponding aggregation node placements. \\
	
		\begin{footnotesize}
	\noindent{Inputs:}
	\begin{itemize}
		\item $G(V,E)$: A graph that models the fiber-optic network, consisting of a set of nodes $V$ and optical fiber links $E$ . Each link $e \in E$ is defined as an ordered pair ($s(e)$, $r(e)$), indicating its source and destination nodes, respectively. \\
		
		\item $D$: A set of traffic demands, where each demand $d \in D$ is characterized by its source node $s(d)$ and destination node $r(d)$. All demands are assumed to require the same traffic volume, equivalent to the capacity of a single wavelength (e.g., 400 Gbps).\\
	\end{itemize}
\end{footnotesize}
\begin{footnotesize}
	\noindent{Variables:}
	\begin{itemize}
		\item $x_{e}^{d}  \in \{0,1\} $: A binary variable set to 1 if demand $d$ is routed through link $e$; 0 otherwise. \\
		\item $z_{e}^{d, v} \in \{0,1\} $: A binary variable equal to 1 if demand $d$ is aggregated at node $v$ with another demand, and the resulting aggregated lightpath uses link $e$; 0 otherwise.\\
		\item $\theta_{v}^{d} \in \{0, 1\} $: A binary indicator that is 1 if demand $d$ is aggregated with another demand at node $v$; 0 otherwise. \\
		
		\item $f_{d_1}^{d_2} \in \{0, 1\} $: A binary variable set to 1 if demands $d_1$ and $d_2$ are aggregated with each other; 0 otherwise. 
		\\
		
	\end{itemize}
\end{footnotesize}
\begin{footnotesize}
	\noindent{Objective Function: Minimize the wavelength-link usage}
	
	\begin{equation} \label{eq:obj1}
		\sum_{d \in D} \sum_{e \in E} x_{e}^{d} - \sum_{d \in D} \sum_{e \in E} \sum_{v \in V}  \frac{z_{e}^{d, v}}{2} \\
	\end{equation}
\end{footnotesize}
\begin{footnotesize}

	\noindent{subject to the following constraints:}
	
	\begin{equation} \label{eq:c3} 
		\begin{split}
			\sum_{e \in {E}: v \equiv s(e)} x_{e}^{d} -\sum_{e \in {E}: v \equiv r(e)} x_{e}^{d}= \\		
			\begin{cases} 
				1 &\mbox{if } v \equiv s(d) \\ 
				-1 & \mbox{if } v \equiv r(d)\\
				$0$ & otherwise \\
			\end{cases}     \qquad \qquad \forall d \in D, \forall v \in V \hfill
		\end{split}
	\end{equation}
	
	\begin{align} \label{eq:c6}  {
			\sum_{v \in V} \theta_{v}^{d} \leq 1 \qquad and \qquad \theta_{v}^{d} = 0 \qquad \mbox{if } v \equiv r(d) \qquad \forall  d \in D
		}
	\end{align}
	
	\begin{align} \label{eq:c7} {
			\sum_{d_2 \in D} f_{d_1}^{d_2} \leq 1 \qquad \forall d_1 \in D
		}
	\end{align}
	
	\begin{equation} \label{eq:c8}
		\sum_{d_2 \in D: r(d_2) \neq r(d_1)}  {f^{d_1}_{d_2}} = 0 \qquad \forall d_1 \in D
	\end{equation}

	\begin{align} \label{eq:c9} {
			f_{d_1}^{d_2} = f_{d_2}^{d_1} \qquad \forall d_1, d_2 \in D
		}
	\end{align}
	
	\begin{align} \label{eq:c10} {
			\sum_{v \in V} z_{e}^{d_1, v} \leq \sum_{d_2 \in D} f_{d_1}^{d_2} \qquad \forall d_1 \in D, \forall e \in E
		}
	\end{align}
	
	\begin{align} \label{eq:c11} {
			\sum_{d_2 \in D} f_{d_1}^{d_2} = \sum_{v \in V} \theta_{v}^{d_1} \qquad \forall d_1 \in D
		}
	\end{align}

	\begin{align} \label{eq:c14} {
			\theta_{v}^{d_1} - \theta_{v}^{d_2}+f_{d_1}^{d_2} \leq 1 \qquad \forall d_1, d_2 \in D, \forall v \in V
		}
	\end{align}
	
	\begin{align} \label{eq:c15} {
			\theta_{v}^{d_2} - \theta_{v}^{d_1}+f_{d_1}^{d_2} \leq 1 \qquad \forall d_1, d_2 \in D, \forall v \in V
		}
	\end{align}
	
	\begin{align} \label{eq:c16} {
			z_{e}^{d, v}  \leq x_{e}^{d}  \qquad \forall d \in D, \forall v \in V, \forall e \in E
		}
	\end{align}
	

\begin{equation} \label{eq:c17}
	\begin{split}
		\sum_{e \in E: i \equiv s(e)} z_{e}^{d, v} - \sum_{e \in E: i \equiv r(e)} z_{e}^{d, v}= \\
		\begin{cases} 
			\theta_{v}^{d} &\mbox{if } i \equiv v \\ 
			-\theta_{v}^{d} & \mbox{if } i \equiv r(d)\\
			0 & \mbox{otherwise} \\
		\end{cases} \qquad \qquad \forall d \in D, \forall v \in V, \forall i \in V \hfill
	\end{split}
\end{equation}

\end{footnotesize}
	
Equation \ref{eq:obj1} defines the objective of minimizing the total wavelength link cost while accommodating all traffic demands. When an aggregation operation is applied, it results in a saving of wavelength links—hence the inclusion of the second term in Eq. \ref{eq:obj1}. The standard flow conservation constraints for each demand is presented in Eq. \ref{eq:c3}.
Constraints in Eq. \ref{eq:c6} ensure that each demand can be aggregated at most once at a selected aggregation node, which must differ from the destination node. To guarantee that a demand is aggregated with at most one other demand sharing the same destination, constraints formulated by Eq. \ref{eq:c7}, Eq. \ref{eq:c8}, and Eq. \ref{eq:c9} are imposed. Constraints in Eq. \ref{eq:c10} and Eq. \ref{eq:c11} maintain consistency between aggregated demands, their associated aggregation nodes, and the aggregation links. Additionally, constraints represented by Eq. \ref{eq:c14} and Eq. \ref{eq:c15} enforce that if two demands are aggregated, the aggregation must occur at the same node. Constraints in Eq. \ref{eq:c16} determine the routing of the resulting aggregated lightpaths. Lastly, constraints in Eq. \ref{eq:c17} ensure flow conservation for the aggregated traffic. \\

The proposed model is formulated as an integer linear programming (ILP) problem, a class of problems recognized for its NP-hard computational complexity. Beyond the conventional decision variables and constraints used to capture routing paths for individual demands, this model introduces additional variables and constraints to handle the interaction between transitional lightpaths for optical computing. In particular, the inclusion of the binary variable $z_{e}^{d, v} \in \{0,1\}$ and the constraints defined in Eqs. \ref{eq:c14}, \ref{eq:c15}, \ref{eq:c16}, and \ref{eq:c17} significantly increase the complexity of the model. As a result, this formulation becomes computationally more challenging by an order of magnitude than the standard routing typically used in optical-bypass networks. \\

\section{Numerical Results}

\begin{figure}[!ht]
	\centering
	\includegraphics[width=\linewidth, height = 5.5cm]{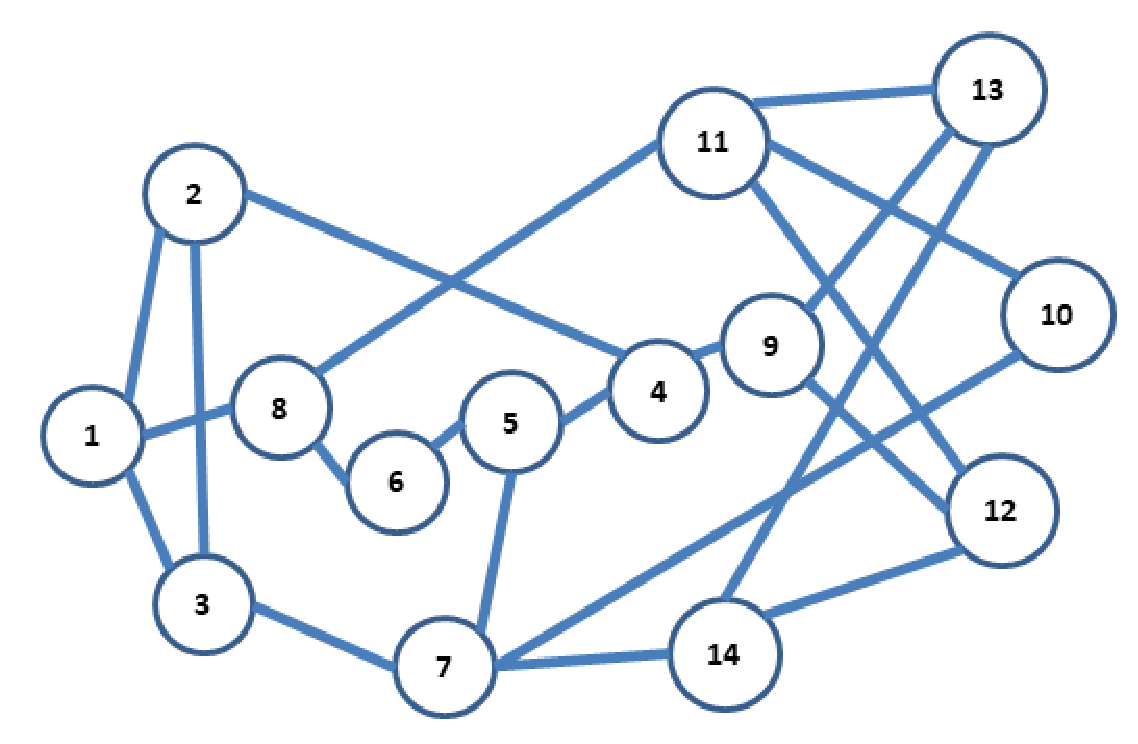}
	\caption{Network Topology under Test (NSFNET)}
	\label{fig:cost239}
\end{figure}

This section presents a series of numerical experiments evaluating the performance of our proposed architecture, which incorporates optical aggregation operation within the optical computing-communication integrated network, in comparison to the traditional optical-bypass model. The study is conducted using the NSFNET topology depicted in Fig. \ref{fig:cost239}. To create favorable conditions for aggregation, traffic demands are generated using a two-to-many scheme: two nodes are randomly designated as sources, while several other nodes serve as randomly selected destinations. This setup naturally increases the probability of demands sharing a common destination\textemdash an essential condition for optical aggregation. \\

We evaluate three traffic load scenarios involving 4, 8, and 12 destination nodes, respectively. For each scenario, 10 distinct traffic samples are simulated. Each demand is provisioned with a dedicated wavelength channel operating in the standard C-band, with 50 GHz spacing between channels. The primary performance metric is the total wavelength link cost required to route all demands and all the results are optimally collected from solving the proposed ILP model in Sect. III for a fair comparative evaluation. \\

The results in Fig. 4 through Fig. 6 report the wavelength-link usage for both designs in supporting the three increasing traffic loads, respectively and it is clearly observed that the optical-aggregation-enabled design is spectrally more efficient than the optical-bypass one in all simulated cases, with the maximum relative gain reaching slightly more than $35\%$.

\begin{figure}[!ht]
	\centering
	\includegraphics[width=\linewidth, height = 5cm]{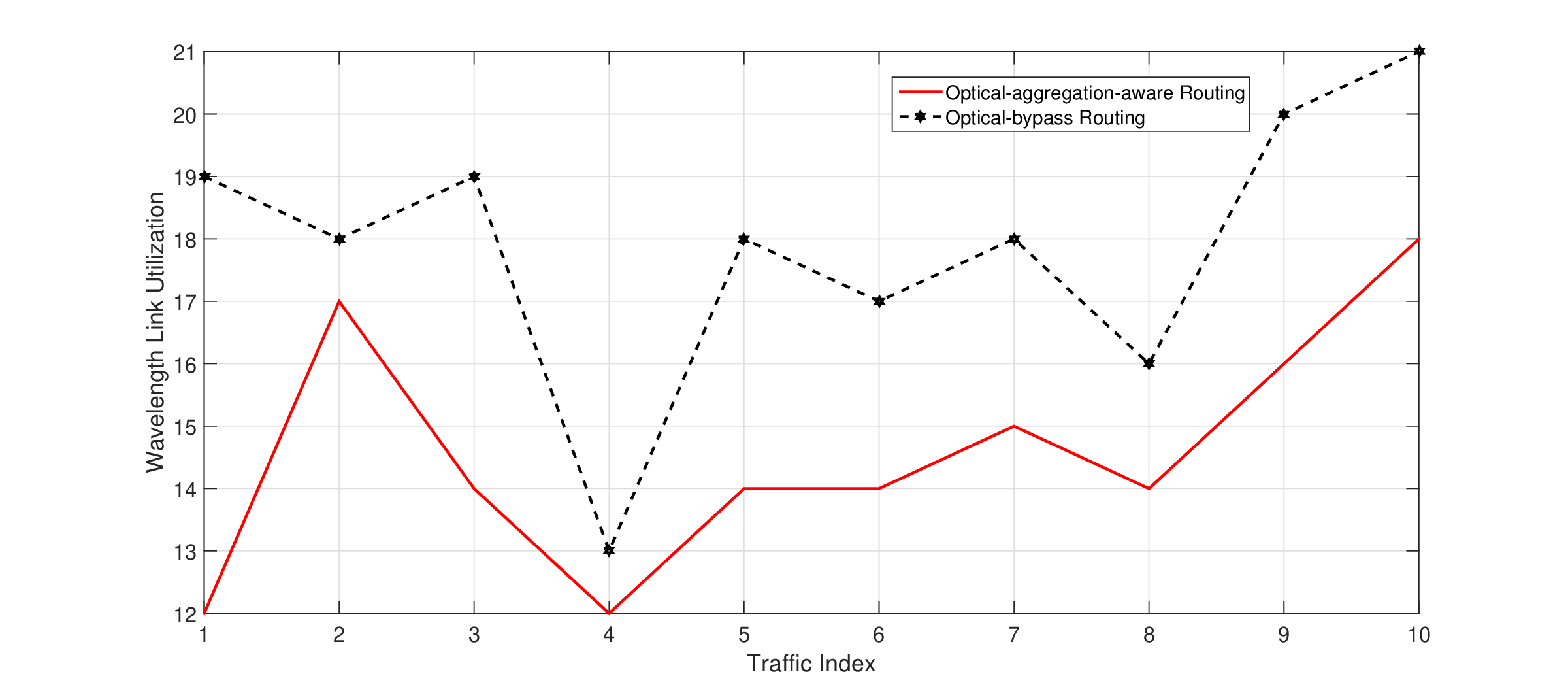}
	\caption{Comparative routing cost in low-load scenarios}
	\label{fig:i5}
\end{figure}

\begin{figure}[!ht]
	\centering
	\includegraphics[width=\linewidth, height = 5cm]{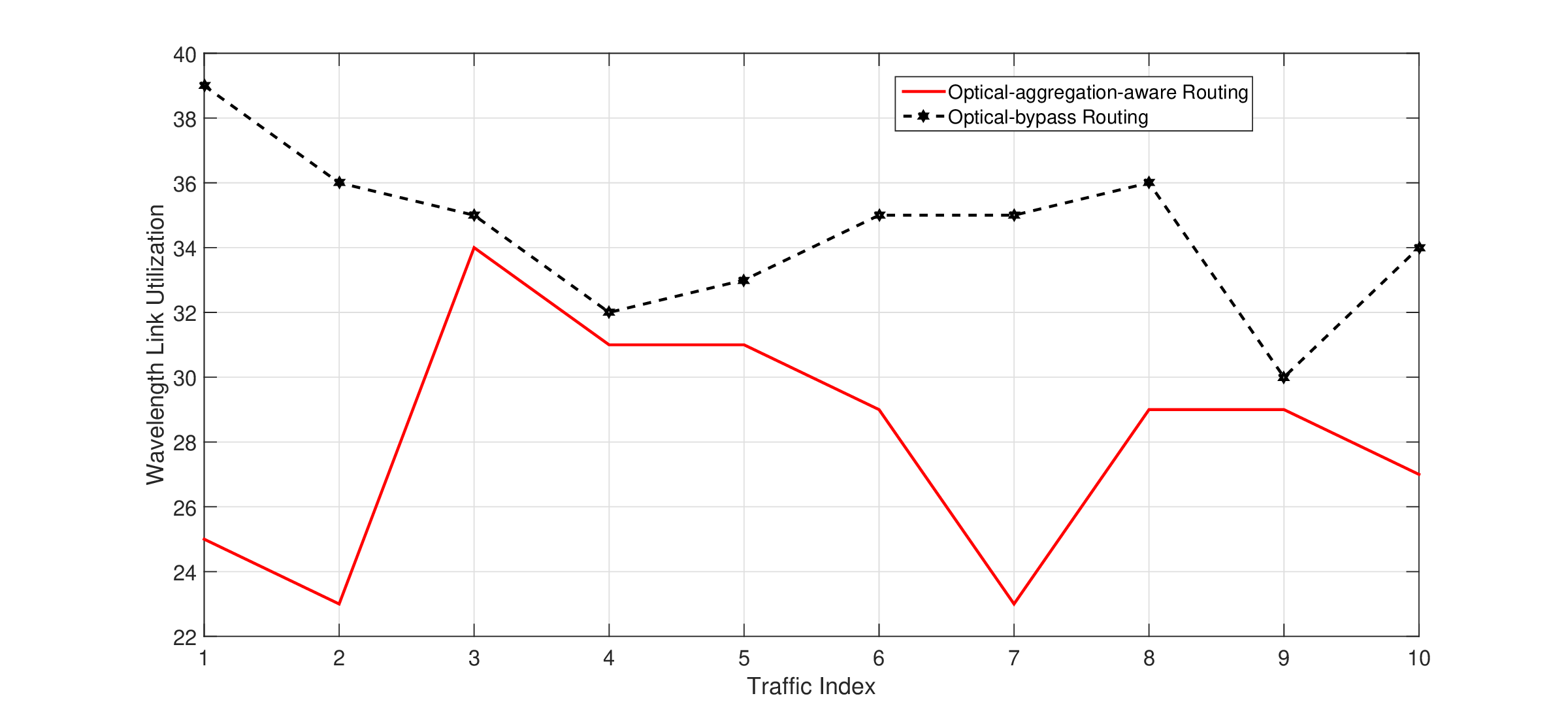}
	\caption{Comparative routing cost in medium-load scenarios}
	\label{fig:i6}
\end{figure}

\begin{figure}[!ht]
	\centering
	\includegraphics[width=\linewidth, height = 5cm]{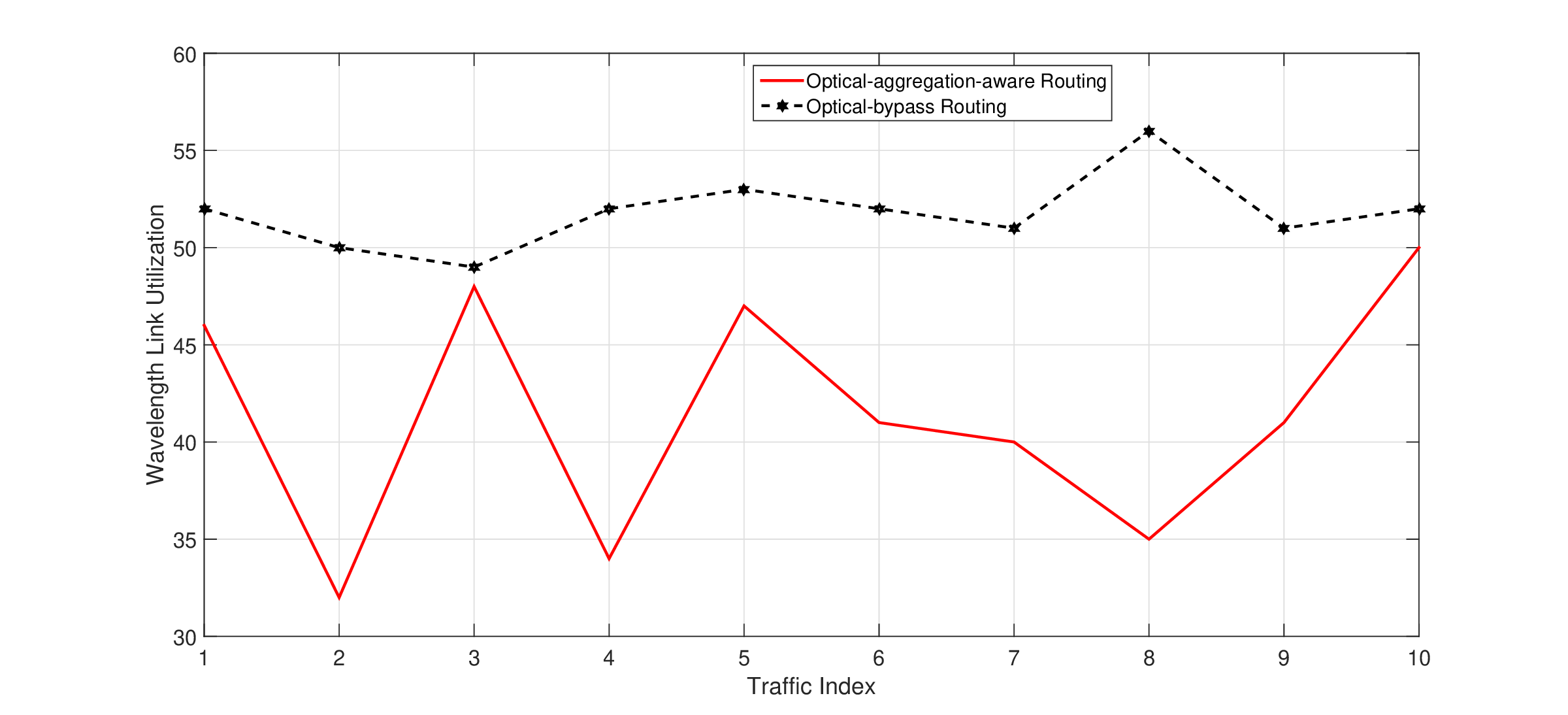}
	\caption{Comparative routing cost in high-load scenarios}
	\label{fig:i7}
\end{figure}

To examine the underlying factors contributing to these gains, a deeper inspection on the optical-aggregation design is shown in Tab. 1 that provides the comprehensive simulation results for one traffic sample of medium-load scenario. Here we would like to draw the attention to the routing of demand $11\rightarrow1$ and $14\rightarrow1$ as an important observation would be uncovered. As shown in Tab. 2, both two demands select their shortest path to minimize the wavelength-link cost in the traditional optical-bypass mode. However, in the aggregation-aware routing shown in Tab. 1, the demand $14\rightarrow1$ is routed over a longer path consisting of four links to make it favorably aggregated to the demand $11\rightarrow1$ and this is justifiable as the spectral benefit derived from the optical aggregation mechanism is higher than the cost of deviation from the shortest path. Specifically, the wavelength-link usage for accommodating two demands $11\rightarrow1$ and $14\rightarrow1$ in optical-aggregation mode is four units, saving one unit compared to the shortest approach in the optical-bypass architecture. 

\begin{table*}[ht]
	\caption{Optical-aggregation-enabled routing}
	\label{tab: r1}
	\centering
	\begin{tabular}{|c|c|c|c|c|}
		\hline
		Traffic Demand & Routing Link(s) & Aggregation Node & Aggregation Link(s) & With Demand  \\
		\hline 
		11$\rightarrow$12 & (11-12) & N/A & N/A &  N/A \\
		11$\rightarrow$10 & (11-10) & N/A & N/A &  N/A \\
		11$\rightarrow$1 & (11-8-1) & 11 & (11-8-1) & 14$\rightarrow$1  \\
		11$\rightarrow$7 & (11-10-7) & N/A & N/A & N/A \\

		14$\rightarrow$12 & (14-12) & N/A & N/A & N/A \\
		14$\rightarrow$10 & (14-7-10) & N/A & N/A & N/A \\ 
		14$\rightarrow$1 & (14-12-11-8-1) & 11 & (11-8-1) & 11$\rightarrow$1 \\
		14$\rightarrow$7 & (14-7) & N/A & N/A & N/A \\

		\hline
	\end{tabular}
\end{table*}

\begin{table*}[ht]
	\caption{Non-aggregation routing in optical-bypass networks}
	\label{tab: r2}
	\centering
	\begin{tabular}{|c|c|}
		\hline
		Demand & Routing \\
		\hline 
		11$\rightarrow$12 & (11-12) \\
		11$\rightarrow$10 & (11-10) \\
		11$\rightarrow$1 & (11-8-1) \\
		11$\rightarrow$7 & (11-10-7) \\
		
		14$\rightarrow$12 & (14-12)  \\
		14$\rightarrow$10 & (14-7-10)  \\ 
		14$\rightarrow$1 & (14-7-3-1) \\
		14$\rightarrow$7 & (14-7)  \\
		
		\hline
	\end{tabular}
\end{table*}

\section{Conclusion}

This paper presents a visionary perspective for future optical transport networks wherein optical computing operations could be performed between lightpath entities being routed over the same intermediate nodes, possibly in a backward-compatible manner. A case study focusing on optical aggregation, a special type of computing between lightpaths of the same format and line-rate, has been examined, in reference to the traditional optical-bypass architecture. A mathematical formulation for optical-aggregation-aware network has been developed and numerically evaluated on the realistic NSFNET topology revealing the spectral efficiency enhancement compared to the optical-bypass mode. \\

This study also underscores the emerging capability of fiber-optic networks to serve not only as high-capacity communication channels but also as platforms for in-network computation\textemdash paving the way toward realizing optical-layer intelligence. This paradigm shift challenges the traditional notion of optical systems as mere data conduits and positions them as active participants in processing tasks. The growing demand for training large-scale foundation models, coupled with increased investment in optical computing technologies to address limitations in electronic processing, highlights a critical need. Moving forward, the development of integrated optimization frameworks that jointly manage both computation and communication resources within the optical domain presents a compelling direction for future research. With the growing maturing of optical computing platforms, we envision the extended applicability of this architecture, from long-haul to access and optical data center networks.

\bibliographystyle{IEEEtran}
\bibliography{IEEEabrv,ref}

\vspace{12pt}
\color{red}

\end{document}